\documentclass[aps,prl,superscriptaddress,twocolumn]{revtex4-1}

\usepackage[utf8]{inputenc}
\usepackage{amsmath} 
\usepackage{amssymb} 
\usepackage{amsfonts} 
\usepackage{graphicx}   
\usepackage{verbatim}  
\usepackage{color}     
\usepackage{hyperref}   
\usepackage{natbib}
\usepackage{gensymb}
\usepackage{color}
\usepackage{multirow}
\usepackage{xfrac}

\begin{document}

\title{Unveiling the Semimetallic Nature of 1$T$-TiSe$_2$ by Doping its Charge Density Wave}

\author{T. Jaouen}
\altaffiliation{Corresponding author.\\ thomas.jaouen@unifr.ch}
\affiliation{D{\'e}partement de Physique and Fribourg Center for Nanomaterials, Universit{\'e} de Fribourg, CH-1700 Fribourg, Switzerland}

\author{M. Rumo}
\affiliation{D{\'e}partement de Physique and Fribourg Center for Nanomaterials, Universit{\'e} de Fribourg, CH-1700 Fribourg, Switzerland}

\author{B. Hildebrand}
\affiliation{D{\'e}partement de Physique and Fribourg Center for Nanomaterials, Universit{\'e} de Fribourg, CH-1700 Fribourg, Switzerland}

\author{M.-L. Mottas}
\affiliation{D{\'e}partement de Physique and Fribourg Center for Nanomaterials, Universit{\'e} de Fribourg, CH-1700 Fribourg, Switzerland}

\author{C. W. Nicholson}
\affiliation{D{\'e}partement de Physique and Fribourg Center for Nanomaterials, Universit{\'e} de Fribourg, CH-1700 Fribourg, Switzerland}

\author{G. Kremer}
\affiliation{D{\'e}partement de Physique and Fribourg Center for Nanomaterials, Universit{\'e} de Fribourg, CH-1700 Fribourg, Switzerland}

\author{B. Salzmann}
\affiliation{D{\'e}partement de Physique and Fribourg Center for Nanomaterials, Universit{\'e} de Fribourg, CH-1700 Fribourg, Switzerland}

\author{F. Vanini}
\affiliation{D{\'e}partement de Physique and Fribourg Center for Nanomaterials, Universit{\'e} de Fribourg, CH-1700 Fribourg, Switzerland}

\author{C. Barreteau}
\affiliation{Department of Quantum Matter Physics, University of Geneva, 24 Quai Ernest-Ansermet, 1211 Geneva 4, Switzerland}

\author{E. Giannini}
\affiliation{Department of Quantum Matter Physics, University of Geneva, 24 Quai Ernest-Ansermet, 1211 Geneva 4, Switzerland}

\author{H. Beck}
\affiliation{D{\'e}partement de Physique and Fribourg Center for Nanomaterials, Universit{\'e} de Fribourg, CH-1700 Fribourg, Switzerland}

\author{P. Aebi}
\affiliation{D{\'e}partement de Physique and Fribourg Center for Nanomaterials, Universit{\'e} de Fribourg, CH-1700 Fribourg, Switzerland}

\author{C. Monney}
\affiliation{D{\'e}partement de Physique and Fribourg Center for Nanomaterials, Universit{\'e} de Fribourg, CH-1700 Fribourg, Switzerland}

\begin{abstract}

The semimetallic or semiconducting nature of the transition metal dichalcogenide 1$T$-TiSe$_2$ remains under debate after many decades mainly due to the fluctuating nature of its 2 $\times$ 2 $\times$ 2 charge-density-wave (CDW) phase at room-temperature. In this letter, using angle-resolved photoemission spectroscopy, we unambiguously demonstrate that the 1$T$-TiSe$_2$ normal state is semimetallic with an electron-hole band overlap of $\sim$110 meV by probing the low-energy electronic states of the perturbed CDW phase strongly doped by alkali atoms. Our study not only closes a long-standing debate but also supports the central role of the Fermi surface for driving the CDW and superconducting instabilities in 1$T$-TiSe$_2$.

\end{abstract}

\date{\today}
\maketitle

Layered transition-metal dichalcogenides (TMDCs) have attracted considerable interest due to their rich phase diagrams including intertwined charge-density-waves (CDWs), Mott states and superconductivity \cite{Morosan2006a, Sipos2008, Kusmartseva2009a}.
A common feature that explains the intrinsic tendency of these materials to exhibit electronic phase transitions is that their density of states close to the Fermi level ($E_F$), dominated by the $d$-orbitals of the transition metal layers, is usually high \cite{Novoselov2016, Wilson1969}. Although 1$T$-TiSe$_2$ is a prototypical material with a 2 $\times$ 2 $\times$ 2 commensurate CDW occuring at $\sim$200 K and hosting superconductivity under pressure, Cu-doping or electrical gating \cite{Kusmartseva2009a, Morosan2006a, Li2015}, its normal state still remains elusive after many decades, due to the fluctuating nature of the CDW at room-temperature (RT) \cite{Jaouen2018}. Whereas transport measurements \cite{salvo1976, Wilson1978b} as well as recent optical spectroscopy and conductivity studies \cite{Li2007, Velebit2016}, all concluded on a semimetallic normal state with electron-hole ($e-h$) band overlap $\sim$-100 meV, three-dimensional (3D) momentum-resolved angle-resolved photoemission spectroscopy (ARPES) measurements have reported RT values of band gaps up to 150 meV \cite{Rossnagel2002b, Kidd2002, Zhao2007, Rasch2008, Chen2016, Watson2019}. 

\begin{figure}[h!]
\includegraphics[width=0.47\textwidth]{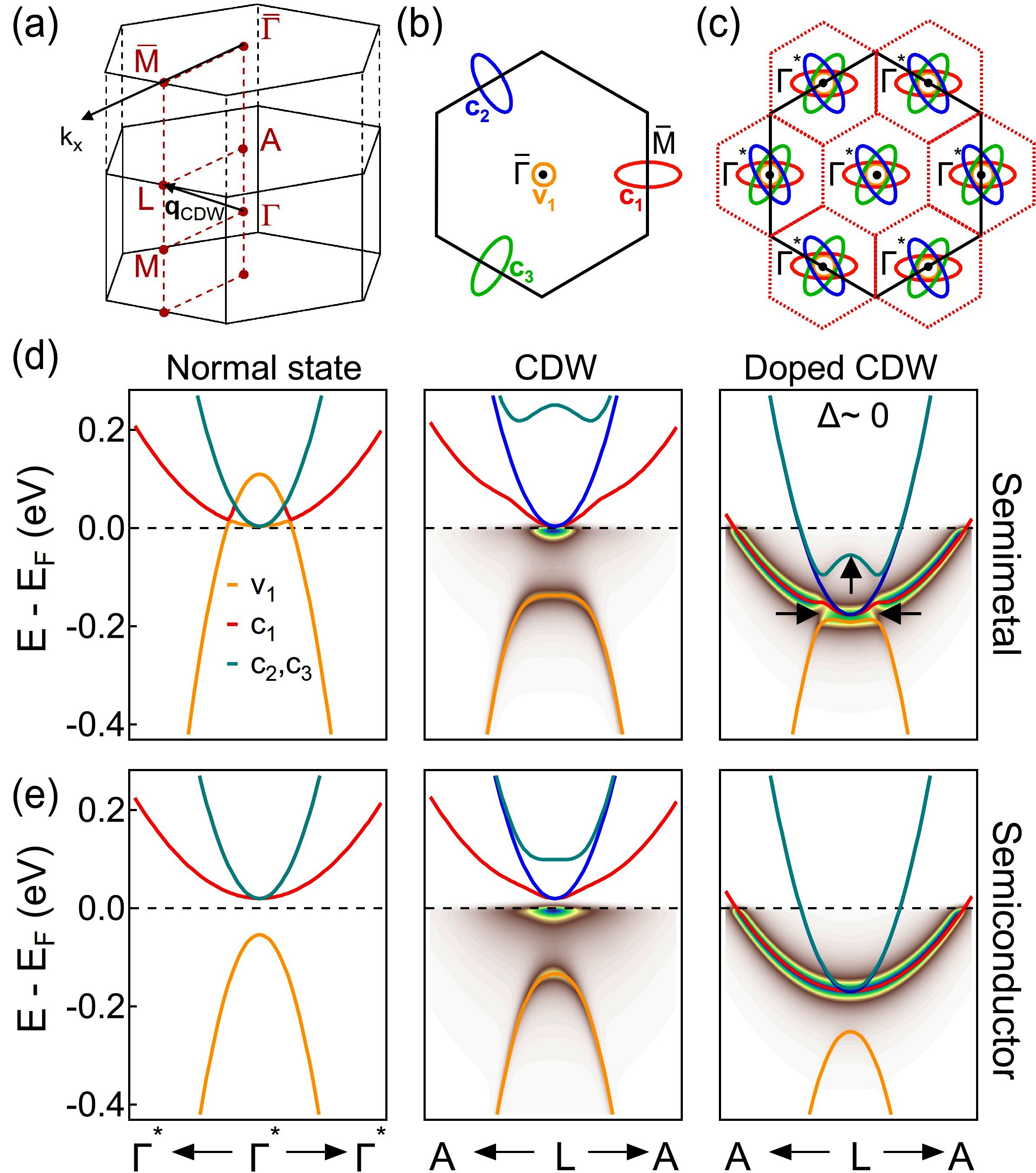}
\caption{(a) 3D BZ of 1$T$-TiSe$_2$. $\textbf{q}_{CDW}$, connects $\Gamma$ and one of the $L$ points at the CDW phase transition. The $k_x$ axis depicted by the black arrow is the $A$-$L$ direction of the bulk BZ. (b)-(c) Schematic pictures of the FS of 1$T$-TiSe$_2$ in the normal state ($\Delta =$ 0 meV) and in the CDW phase ($\Delta \ne$ 0 meV). In panel (b) the FS is composed of the valence band $v_1$ at $\Gamma$ and three conductions bands $c_1$, $c_2$, and $c_3$ at $L$. In panel (c), the electron pockets at $L$, backfolded to $\Gamma$ produce "flowerlike" FSs at each new high-symmetry point $\Gamma^*$. (d)-(e) Left panel: calculated dispersions of the semimetallic and semiconducting normal states along $\Gamma^*$-$\Gamma^*$. Middle and right panels: calculated dispersions and spectral functions along $A$-$L$ for the undoped ($\Delta =$ 118 and 45 meV in (d) and (e), respectively) and doped ($\Delta \sim$ 0 meV) CDW states.}\label{fig1}
\end{figure}

The key point is that the bare electronic band structure of the 1$T$-TiSe$_2$ non-distorted phase is not accessible to RT ARPES measurements due to the presence of strong short-range fluctuations. The result is that ARPES probes a "pseudogap" CDW state as manifested by the presence of diffuse backfolded band intensities and suppression of spectral weight \cite{Jaouen2018, Monney2015, Monney2012b, Borisenko2008, Pillo1999}. In this letter, we show that the normal-state band structure of 1$T$-TiSe$_2$ can be revealed from the low-temperature (LT) CDW state strongly doped by alkali atoms. This therefore closes a long-standing debate by demonstrating that 1$T$-TiSe$_2$ is a semimetal with an $e-h$ band overlap of $\sim$110 meV.

In the 1 $\times$ 1 $\times$ 1 normal state, the 1$T$-TiSe$_2$ low-energy electronic states consist of a Se 4$p$ hole pocket (labeled $v_1$) at $\mathrm{\Gamma}$ and Ti 3$d$ electron pockets (labeled $c_1$, $c_2$, and $c_3$) at the three equivalent $L$ points of the 3D Brillouin zone (BZ) [see Fig. \ref{fig1} (a)-(b)]. At the CDW transition, i.e. with a non-zero order parameter $\Delta$, the $\mathrm{\Gamma}$ and the $L$ points are connected by the three new reciprocal lattice $q$-vectors leading to the doubling of the lattice periodicity [Fig. \ref{fig1}(c)]. As a result, $\Gamma$ becomes equivalent to $L$ and the electron pockets at $L$, backfolded to $\Gamma$ produce "flowerlike" Fermi surfaces (FSs) at each newly equivalent high-symmetry point $\Gamma^*$ \cite{Monney2012}.  

\begin{figure}[b]
\includegraphics[width=0.45\textwidth]{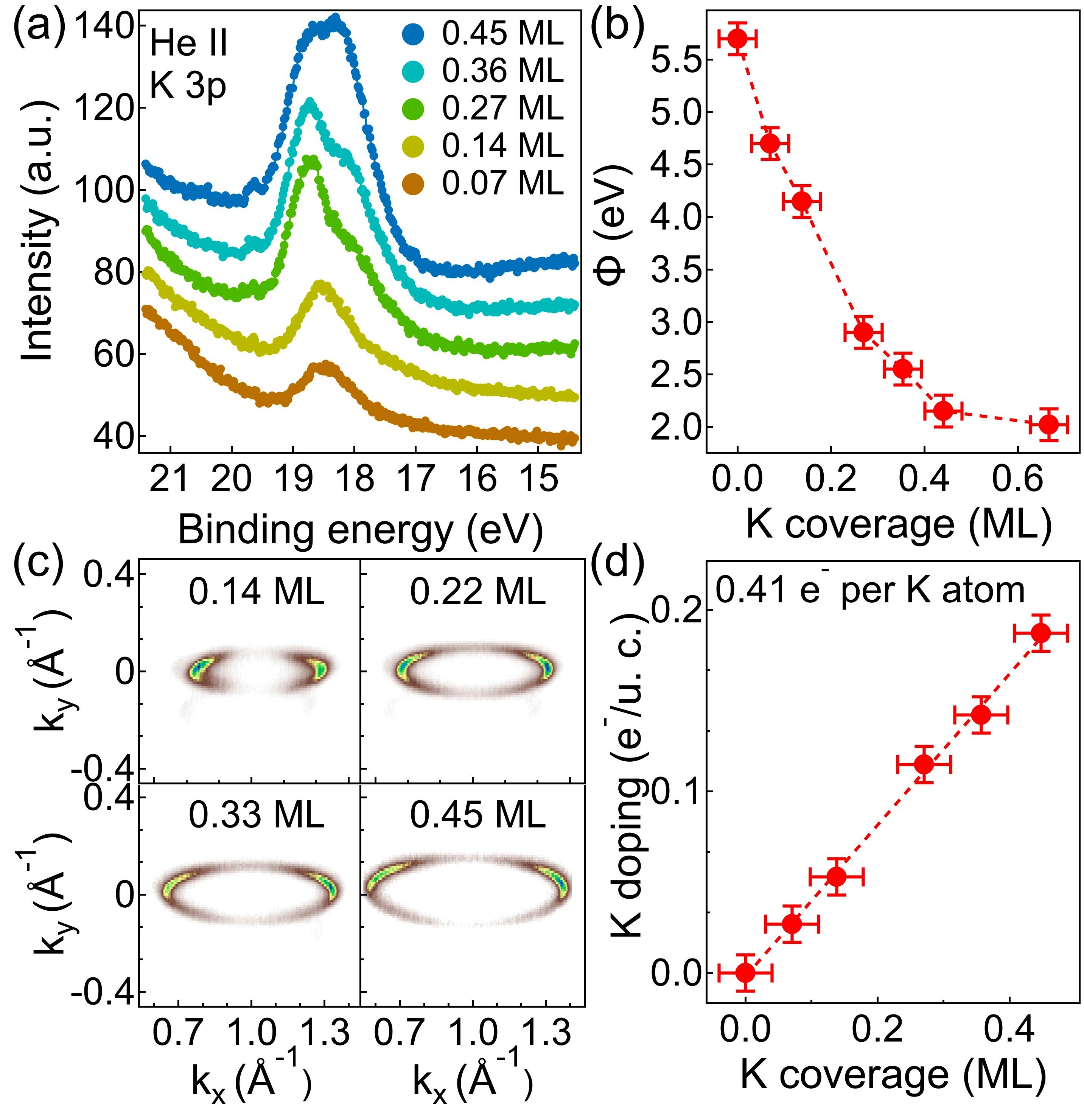}
\caption{(a) Evolution of the K 3$p$ shallow core level upon sequential K evaporations at T$=$40 K on 1$T$-TiSe$_2$ measured using He-II photon energy ($h\nu$=$40.8$ eV). (b) 1$T$-TiSe$_2$ work function changes as a function of the K coverage. (c) Evolution of the 1$T$-TiSe$_2$ Fermi surface at $L$ measured at T$=$30~K as a function of K coverage allowing for determining the surface doping from the Luttinger area increase. (d) Linear dependence of the K surface doping as a function of K coverage. The charge transfer of 0.41 $e-$ per K atom is obtained from the slope of the doping vs. coverage curve.}\label{fig2}
\end{figure}

The most recent proposals for the electron and hole band configurations in the normal state \cite{Jaouen2018, Watson2019}, are an initial semimetallic $e-h$ band overlap of 110 meV \cite{Jaouen2018} [Fig. \ref{fig1} (d), left panel] or a semiconducting band gap of 74 meV \cite{Watson2019} [Fig. \ref{fig1} (e), left panel]. The corresponding calculated band dispersions and spectral functions of the CDW state \footnote{Details on the model from which the near-$E_F$ dispersions and spectral functions describing ARPES have been computed can be found in Refs. \cite{Monney2009a, Jaouen2018}.} [Fig. \ref{fig1} (d)-(e), middle panels] for both cases are very similar, despite having different values of $\Delta$ (118 and 45 meV for the semimetal and the semiconductor, respectively) calculated for obtaining identical values of the binding energy (BE) of the backfolded band $v_1$ \footnote{In a photoemission experiment, we can extract the order parameter $\Delta$ from the shift of the backfolded valence band, ${\Delta}E_{backf}$ using,
\begin{eqnarray}
\Delta=\left(\frac{\Delta E_{backf}(\Delta E_{backf}-E_g)}{3}\right)^\frac{1}{2} 
\end{eqnarray}
where $E_g$ is the $e-h$ band overlap ($E_g$ $<$0) or band gap ($E_g$ $>$0) in the normal state.\cite{Jaouen2018}}. Indeed, compared with both of the normal state band configurations, in the CDW state the valence band $v_1$ [orange band] and the conduction band $c_3$ [green band] repel each other symmetrically around $E_F$. The $c_1$ and $c_2$ bands remain at their original positions and in addition the $c_2$ band [blue band, middle panels] has no spectral weight \cite{Monney2009a}. Therefore, the normal state of 1$T$-TiSe$_2$ can not be extrapolated from photoemission measurements of the CDW state. 

However, upon sufficient electron-doping which induces a chemical-potential shift ($\Delta \mu$) and a reduction of $\Delta$, clear differences between the LT band structures of an initial semimetallic and semiconducting band configurations now appear [Fig. \ref{fig1}(d)-(e), right panels]. In particular, the "smoking gun" spectral fingerprints in ARPES for a semimetallic normal-state are: the so-called "Mexican hat" shape of $c_3$ [black vertical arrow on the right panel of Fig. \ref{fig1}(d)] associated with the top of the CDW-split hole band shifted below $E_F$; gaps at the avoided crossings of the electron and hole bands [black horizontal arrows on the right panel of Fig. \ref{fig1}(d)]; as well as the shift of the backfolded band ($v_1$) towards $E_F$. For that purpose, here we electron-dope the CDW state of 1$T$-TiSe$_2$ using LT depositions of potassium (K) atoms in order both to access the unoccupied part of the band structure and to lower $\Delta$ which is known to be highly sensitive to doping \cite{Qian2007, Jaouen2018, Hildebrand2017, Li2015}. 

The 1$T$-TiSe$_2$ single crystals were grown by chemical vapor transport at 590 \celsius , therefore containing less than 0.20 $\%$ of native Ti impurities \cite{Hildebrand2016, Hildebrand2014}. Clean surfaces were obtained by cleaving in ultrahigh vacuum at RT. K atoms were then evaporated \textit{in situ} from a carefully outgassed SAES getter source onto the TiSe$_2$ surface kept at $\sim$40 K to inhibit K intercalation \cite{Rossnagel2010, Caragiu2005}. During the K evaporation, the pressure was maintained below 5 $\times$ 10$^{-10}$ mbar. The ARPES measurements were carried out using a Scienta DA$30$ photoelectron analyzer with He-I radiation ($h\nu$=$21.2$ eV) and laser excitation source ($h\nu$=$6.3$ eV) at T $=$ 30 K. The total energy resolution was 5 meV and the base pressure during experiments was better than 1.5 $\times$ 10$^{-10}$ mbar. 

\begin{figure*}[t]
\includegraphics[width=0.7\textwidth]{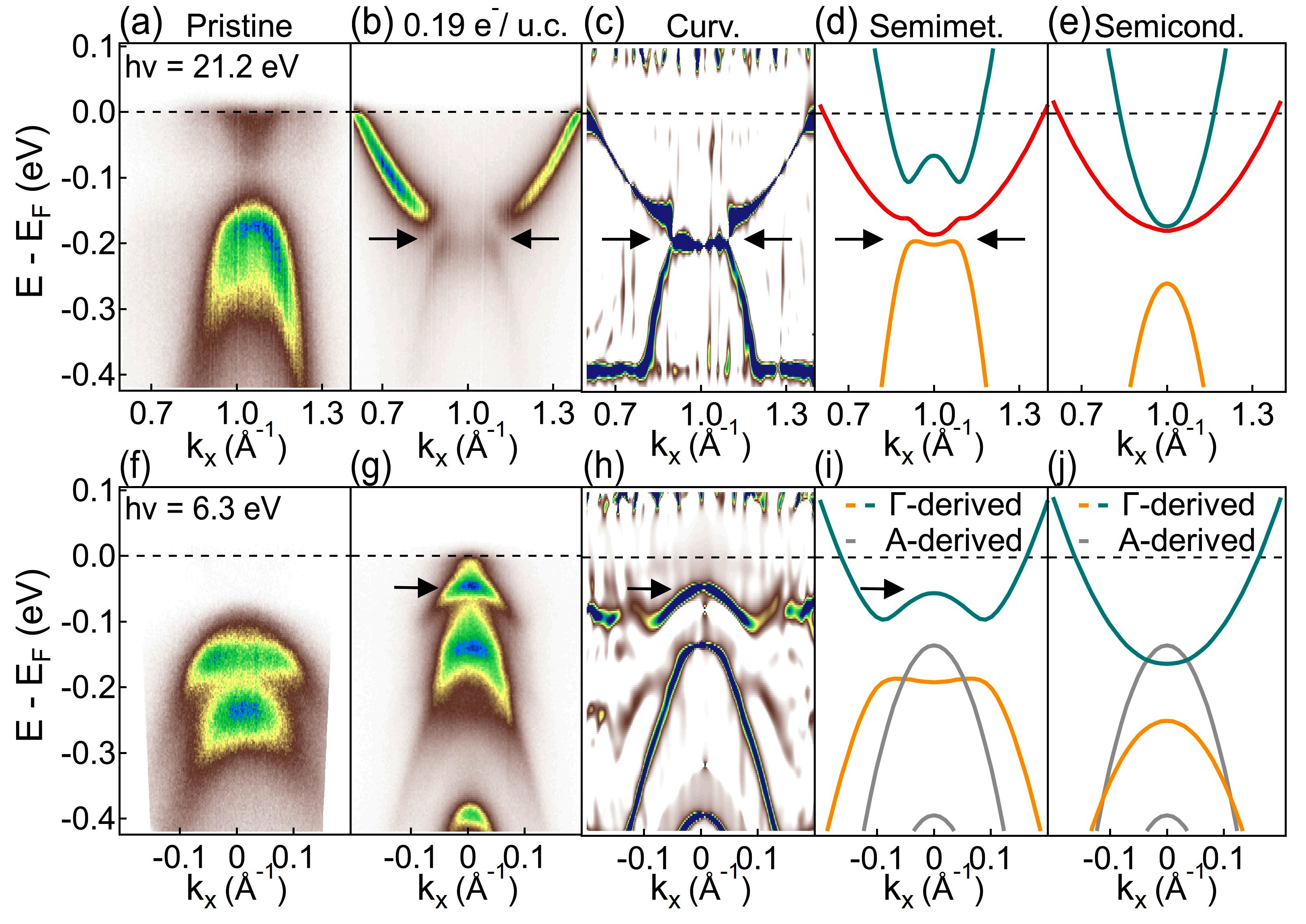}
\caption{(a)-(b) ARPES intensity maps taken at T = 30 K at the $L$ point of the BZ as probed using 21.2 eV photon energy, of pristine (a) and electron-doped (0.19 $e^-$ per unit cell) 1$T$-TiSe$_2$ (b). (c) Curvature plot extracted from the ARPES intensity map of the K-doped 1$T$-TiSe$_2$ surface (b). (d)-(e) Band dispersions for semimetallic (d) and semiconducting (e) band configurations calculated with $\Delta$ values reduced by 80 $\%$ compared to the undoped CDW state and $\Delta \mu$= 180 meV. The black arrows on (b), (c) and (d) show the gap opening at the avoided crossings of the electron and hole bands. (f)-(j) Same than (a)-(e) using 6.3 eV photon energy. The $A$-derived valence bands dispersions on (i) and (j) are not taken into account in the four-band model since it does not include any $k_z$ dependence of the order parameter $\Delta$. They have been obtained from parabolic fits of the experimental momentum distribution curves extracted from the curvature plot in (h) and added to the $\Gamma$-derived calculated ones in (i) and (j). The black arrows on (g), (h), and (i) indicate the Mexican hat.
}\label{fig3}
\end{figure*}

Figure \ref{fig2}(a) shows the evolution of the K 3$p$ shallow core level upon sequential K evaporations at T$\sim$40 K on 1$T$-TiSe$_2$ measured using He-II photon energy ($h\nu$=$40.8$ eV). For K coverages lower than 0.10 ML (1 ML corresponds to one K atom per 1$T$-TiSe$_2$ surface unit cell), the photoemission spectrum mainly consists of one component at 18.4 eV BE associated with 2D dispersed K atoms. With increased coverages, a second component corresponding to closely packed K atoms continuously grows on the low BE side (17.9 eV BE) and starts to be dominant at 0.45 ML coverage as the dispersed alkali atoms begin to condense into metallic islands in close analogy with the mechanism of LT K adsorption on graphite \cite{Caragiu2005}. The work function of 1$T$-TiSe$_2$ changes continuously as a function of the K coverage [Fig. \ref{fig2}(b)] \footnote{The work function, defined as the energy of the vacuum level with respect to $E_F$, is determined from the low-energy cut off of the secondary photoelectron emission. To this end, the samples were biased at -8V to make the measurements of the very low-energy region of the spectrum reliable.}, and exhibits no minimum, but a saturation after the K deposition of 0.45 ML at the K bulk value (1.8 eV). Most of the work function decrease occurs within the first 0.30 ML, indicating a strong electron transfer from the alkali atoms in the 2D dispersed phase to the surface.

The surface doping can be calculated from the Luttinger area of the LT FS at $L$ that expands upon K deposition ($n_{2D} = g_v \frac{k_S k_L}{2\pi}$, where $g_v$ = 3 is the valley multiplicity for all occupied Ti 3$d$ electron pockets, and $k_S$ and k$_L$ are their short and long axes lengths)[Fig. \ref{fig2}(c)]. As shown in Fig. \ref{fig2}(d), the surface electron doping linearly increases with the K coverage and has a value of 0.19 $e^-$ per unit cell at 0.45 ML, which is more than three times the critical doping of 0.06 $e^-$ per unit cell at which the superconductivity emerges in Cu-intercalated 1$T$-TiSe$_2$ \cite{Morosan2006a}, but is similar to that obtained in thin TiSe$_2$ films grown on TiO$_2$ \cite{Jia2018}. In this range of coverage, there is a constant charge transfer from K to the 1$T$-TiSe$_2$ surface of 0.41 $e^-$ per K atom demonstrating that a rigid band model of electron doping applies.

Having demonstrated that LT K depositions effectively strongly electron-dope the 1$T$-TiSe$_2$ surface, let us now focus on the CDW state. Figure \ref{fig3}(a) shows ARPES intensity map of pristine 1$T$-TiSe$_2$ taken at T = 30 K  using He-I radiation ($h\nu$=$21.2$ eV) that probes the initial states close to the $L$ point of the 3D BZ in a free-electron final-state picture \cite{Pillo2000}. We clearly see the well-known backfolding of the topmost Se 4$p_{x,y}$ hole band coming from the $\Gamma$ point as well as the Ti $3d$ electron pocket $c_1$ that derives purely from the $d_{z^2}$ orbital projection of the conduction bands and remains unaffected by the CDW phase transition \cite{Watson2019, Jaouen2018}. Upon strong K doping [0.19 $e^-$ per unit cell, Fig. \ref{fig1}(b)], the Ti 3$d_{z^2}$ band at $L$ rigidly shifts by $\sim$180 meV towards higher BE and starts to \textit{overlap} with the backfolded hole band due to the reduced $\Delta$ (also manifested by the strongly lowered spectral weight of the backfolding), resulting in band gap openings at the avoided $e-h$ band crossings [see black arrows on Fig. \ref{fig3}(b)-(d)]. 

From Fig. \ref{fig1}, the conclusion is clear: the presence of an energy gap at the $e-h$ band crossings and the flattened band dispersion [see the curvature plot, Fig. \ref{fig3}(c), extracted from the ARPES intensity map \cite{Zhang2011}] are only compatible with an initial semimetallic band configuration as exemplified by the direct comparison of the calculated band dispersions of the semimetal and semiconductor cases [Fig. \ref{fig3}(d) and (e), respectively]\footnote{The $c_2$ band is not shown here since it has been demonstrated to carry no spectral weight for every $k$ along the $\Gamma M$ and $AL$ directions within the effective Hamiltonian model of the four interacting bands of 1$T$-TiSe$_2$ \cite{Monney2009a}}. Our calculated band dispersions for the semimetal indeed allow to track both the band gap opening [see black arrows in Fig. \ref{fig3}(d)], and the specific shape and BE of the backfolded hole band [orange band dispersion in Fig. \ref{fig3}(d)]. On the contrary, in the semiconducting case, the CDW band gap cannot be smaller than the initial band gap such that $e$-$h$ band crossings can not occur resulting in a backfolding BE much higher than in experiment.  

In contrast to the backfolded hole band and in agreement with the calculated spectral function [Fig. \ref{fig1}(d)], the Mexican hat [green band dispersion in Fig. \ref{fig3}(d)], hardly distinguishable on the curvature plot [Fig. \ref{fig3}(c)], carries extremely weak spectral weight at $L$. This is a strong indication that the hybridization between the hole band at $\Gamma$ and the electron one at $L$ at the CDW phase transition is accompanied by band inversion. In fact, the Mexican hat corresponds to the top of the original hole band split by the CDW and mainly derives from the Se 4$p$ orbitals, whereas the top of the backfolded hole band is associated with the bottom of the original Ti 3$d$ electron bands. From this, we can explain the strong photoemission intensity of the umklapp of Ti 3$d$ orbital character at $L$ and anticipate that the Se 4$p$-derived Mexican hat should appear more clearly at the $\Gamma$ point since the spectral weight as seen by ARPES also follows the projection of the states onto the non-reconstructed zone \cite{Voit2000}.

Laser-based ARPES measurements ($h\nu$=$6.3$ eV) at normal emission [Fig. \ref{fig3}(f)-(g)] indeed reveal clear changes in the photoemission spectral weight. Whereas for pristine 1$T$-TiSe$_2$ [Fig. \ref{fig3}(f)], one can easily recognize the Se 4$p_{x,y}$ valence bands coming from the $\Gamma$ point with a characteristic flattening of the topmost band, upon K doping [Fig. \ref{fig3}(g)], the Mexican hat, which has shifted below $E_F$, becomes intense [black arrows] and the backfolded band $v_1$ has almost no spectral weight [see Fig. \ref{fig3}(h) and the orange band in Fig. \ref{fig3}(i)]. Interestingly, two new hole bands with significant spectral weight now appear in the photoemission intensity map. From our DFT-calculated band structures \footnote{see Supplemental material}, we can attribute them to the two Se 4$p_{x,y}$ valence bands derived from the $A$ point which can be probed in laser-ARPES due to final-state effect of photoemission beyond the free-electron final state approximation \cite{Lindroos1996}. Indeed, at such low photon energy, it is known that direct transitions to unoccupied Bloch-like states are allowed in a large portion of the BZ along $k_z$ \cite{Xiong2017}.  

In our case, the $A$-derived valence bands allow for accessing the $\Gamma$-$A$ $k_z$ dispersion in the ARPES intensity map of the K-doped CDW state and definitively demonstrate that the 1$T$-TiSe$_2$ normal state is semimetallic. Looking at the semiconducting case in Fig. \ref{fig3}(j), the agreement with the experimental band dispersions is poor, namely the absence of the Mexican hat and the presence of a slight $e$-$h$ band overlap in the $A$-$L$ plane of the BZ [Fig. \ref{fig3}(j)]. Furthermore, our measurements also solve the puzzling observation of Ref.\cite{Watson2019}, namely a decreased gap at LT between the $A$-derived and the unhybridized Ti 3$d_{z^2}$ passenger states at $L$ that determine the band gap in the CDW phase. Starting with a semimetallic normal state, the slightly overlapping bands of the $A$-$L$ plane \textit{also} open a small gap ($\sim$15 meV) through the CDW phase transition.     

\begin{figure}[t]
\includegraphics[width=0.46\textwidth]{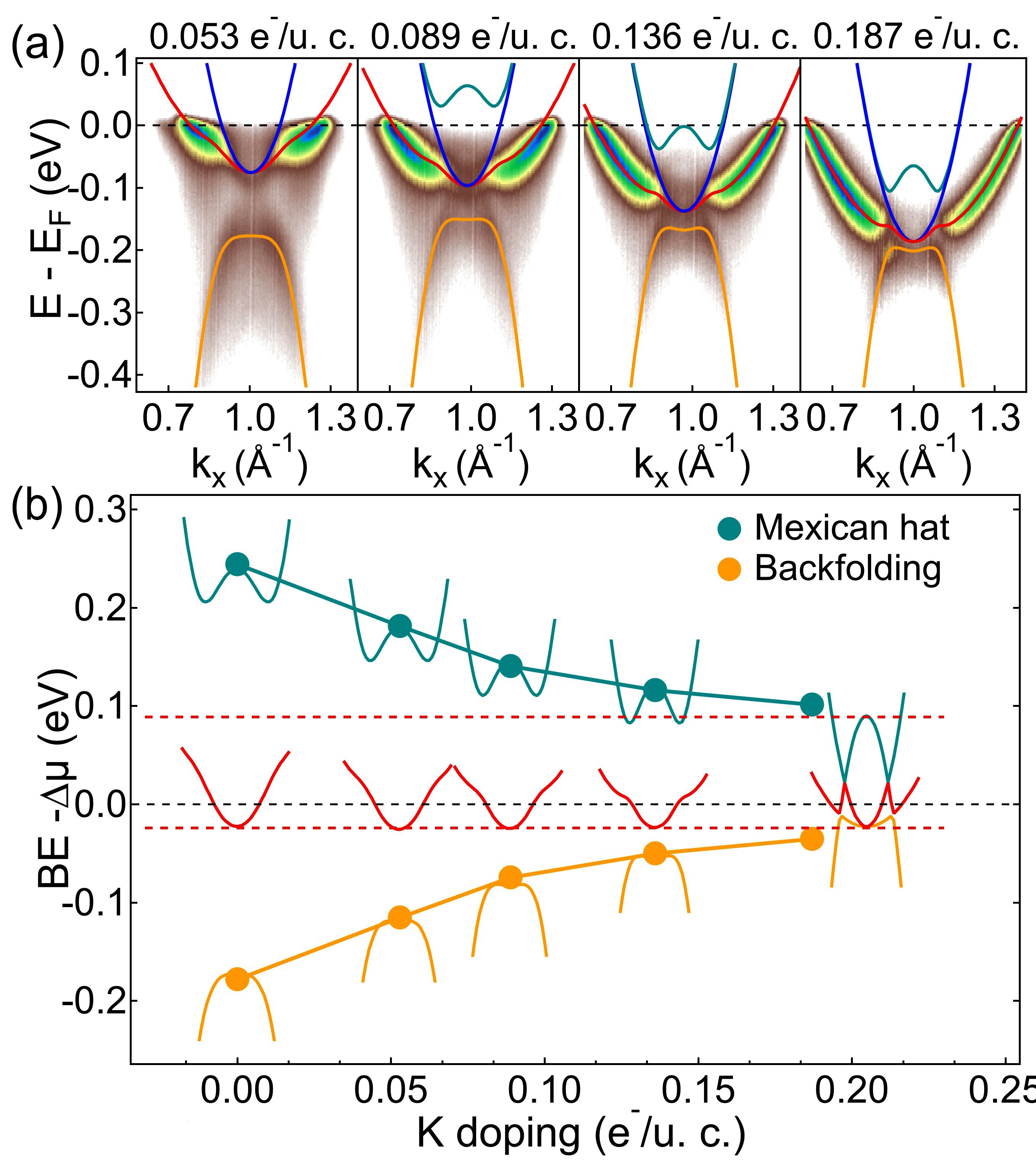}
\caption{(a) ARPES intensity maps at the $L$ point of the BZ upon gradual K doping measured using 21.2 eV photon energy. A logarithmic color scale is used. Calculated near-$E_F$ dispersions are also superimposed. The $\Delta \mu$ values associated with the different K doping concentrations have been extracted from the BE shift of the bottom of the Ti 3$d_{z^2}$ band (red bands) and determined to be 79.5 meV, 100.5 meV, 141.5 meV and  180 meV, from left to right. The $\Delta$ values corresponding to the different K doping concentrations are respectively from left to right, 84 meV, 54 meV, 37 meV and 25 meV. (b) Evolutions of the BEs of the Mexican hat (green bands) and backfolding (orange bands) corrected from $\Delta \mu$ as extracted from the BE shift of the Ti 3$d_{z^2}$ band (red bands) as a function of K doping.}\label{fig4}
\end{figure}

Finally, we show in Fig. \ref{fig4}(a) the ARPES intensity maps at the $L$ point of the BZ upon gradual K doping. The near-$E_F$ band dispersions calculated within our effective Hamiltonian model of four bands in a semimetallic band configuration with an $e-h$ overlap of $\sim$110 meV and interacting through $\Delta$ are also superimposed. Together with the gradual filling of the Ti 3$d_{z^2}$ electron band that linearly shifts with doping (red bands), we can track the CDW gap closure. Figure \ref{fig4}(b) shows the evolution of the BEs of the Mexican hat (green bands) and backfolding (orange bands) corrected by $\Delta \mu$ as extracted from the BE shift of the Ti 3$d_{z^2}$ band (red bands) as a function of K doping. We clearly see that, with increased K doping and decreasing $\Delta$, the Mexican hat and the backfolding band $v_1$ \textit{symmetrically} shift towards the asymptotic limit of vanishing $\Delta$ corresponding to the semimetallic 1$T$-TiSe$_2$ normal state where they respectively correspond to the top and the bottom of the overlapping Se 4$p$ and Ti 3$d$ bands [red-dashed lines on Fig. \ref{fig4}(b)]. 

Our unambiguous demonstration of the semimetallic normal-state of 1$T$-TiSe$_2$ not only closes a long-standing debate, but is also of crucial importance for the understanding of the CDW instability and the closely related emergence of inhomogeneous superconducting states in electron-doped and pressurized compounds. By demonstrating that 1$T$-TiSe$_2$ has a well defined FS in its normal state, our study strongly supports the recently proposed mechanism for both the CDW instability and the emergence of phase-separated states upon doping based on imperfect "nesting" of FS hotspots, i.e. $e-h$ band crossings at the Fermi level \cite{Jaouen2018}. 

\begin{acknowledgments}
This project was supported by the Fonds National Suisse pour la Recherche Scientifique through Div. II and Grant No. P00P2\_170597. Skillful technical assistance was provided by F. Bourqui, B. Hediger and O. Raetzo.
\end{acknowledgments}

\onecolumngrid
\newpage
\begin{center}
{\large\textbf{\boldmath
\textit{Supplemental Material}: Unveiling the Semimetallic Nature of 1$T$-TiSe$_2$ by Doping its Charge Density Wave}}\\[1.5em]

T. Jaouen,$^1$ M. Rumo,$^1$ B. Hildebrand,$^1$ M.-L. Mottas,$^1$ C. W. Nicholson,$^1$ G. Kremer,$^1$\\ B. Salzmann,$^1$ F. Vanini,$^1$ C. Barreteau,$^2$ E. Giannini,$^2$ H. Beck,$^1$ P. aebi,$^1$ and C. Monney$^5$\\[0.5em]

\textit{\small
$^1$D{\'e}partement de Physique and Fribourg Center for Nanomaterials,\\ Universit{\'e} de Fribourg, CH-1700 Fribourg, Switzerland\\
$^2$Department of Quantum Matter Physics, University of Geneva,\\ 
24 Quai Ernest-Ansermet, 1211 Geneva 4, Switzerland\\
}

\vspace{2em}
\end{center}

\twocolumngrid
\setcounter{figure}{0}
\setcounter{equation}{0}
\renewcommand{\thefigure}{S\arabic{figure}}
\renewcommand{\theequation}{S\arabic{equation}}

\subsection{\textit{DFT computational details}}

The DFT-calculated unfolded band structures of the 2 $\times$ 2 $\times$ 2 CDW phase have been obtained within the WIEN2K package \cite{Wien}, using the modified Becke-Johnson (mBJ) exchange potential in combination with local density approximation (LDA) correlation \cite{Tran2009, Koller2012}. Our calculations include spin-orbit coupling (SOC) that has been shown to be crucial for a good description of the Se 4$p$ hole bands along $\Gamma$-$A$ in the TiSe$_2$ normal state \cite{Vydrova2015, Ghafari2017}. The tuning parameter $c$ of the mBJ functional is related to the average value $\bar{g}$ of $g=|\nabla \rho|/\rho$ by $c=A+B\bar{g}^e$ where $\rho$ is the electron density, $A$ and $B$ two free parameters and $e$ an empirical exponent \cite{Koller2012}. In our calculations, $c$ has been fixed for all calculations at 1.34, the value determined to give the best agreement with the measured normal-state band structure of 1$T$-TiSe$_2$ and a minimum of the total energy for atomic displacements corresponding to the periodic lattice distortion (PLD) as proposed by Di Salvo \cite{salvo1976}. The system was modeled using a 2 $\times$ 2 $\times$ 2 superstructure of 8 unit cells of TiSe$_2$ with lattice parameters set to $a$=$b$=3.54 \AA \space and $c$=6.01 \AA \space \cite{salvo1976}. The calculated band structures were unfolded using the FOLD2BLOCH package \cite{Rubel2014}. 

\begin{figure*}[t]
\includegraphics[width=0.65\textwidth]{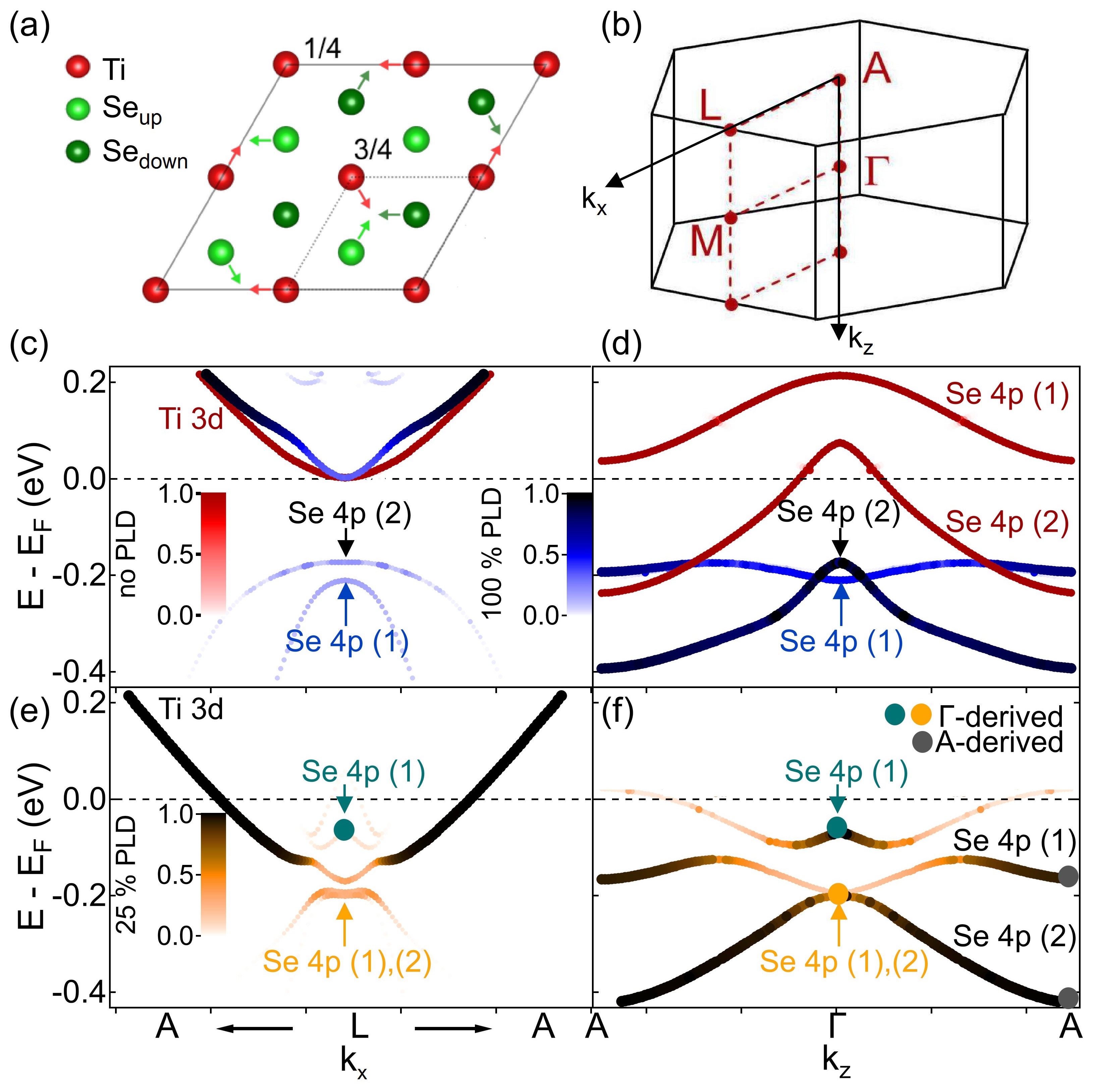}
\caption{(a) Top view (along the $c$-axis) of the PLD accompanying the 1$T$-TiSe$_2$ CDW phase transition (the atomic displacements are represented by arrows). The dotted and full black lines represent the unit cells of the normal phase and of the CDW superstructure, respectively. $\sfrac{3}{4}$ of TiSe$_6$ octahedra are distorted by Ti-Se bond shortening and $\sfrac{1}{4}$ are concerned by a rotation of the Se atoms around a non-moving Ti atom. (b) 3D BZ of 1$T$-TiSe$_2$. The $k_x$ and $k_z$ axes depicted by the black arrows are the $A$-$L$ and $\Gamma$-$A$ directions of the BZ, respectively. (c)-(f) DFT-calculated band structures of 1$T$-TiSe$_2$ along the $ALA$ [(c), (e)] and the $A \Gamma A$ directions [(d), (f)] of the 3D BZ for different magnitudes of the PLD as proposed by Di Salvo \textit{et al.} \cite{salvo1976} (f). (c) and (d) concern the normal state (no PLD, red color scale) and the fully hybridized (100 $\%$ PLD, blue color scale) CDW phase. (e) and (f) respectively focus on the in-plane and out-of-plane DFT band dispersions of a slightly distorted CDW state (25 $\%$ PLD, orange color scale), corresponding to the $\Delta$ reduction  induced by the strongest K-doping of 0.19 $e^-$ per unit cell. The green and orange circles on (f) indicate the maxima of the $\Gamma$-derived bands of the strongly K-doped 1$T$-TiSe$_2$ crystal measured in laser-based ARPES, whereas the grey ones are associated with the maxima of the $A$-derived valence bands.} \label{figS1}
\end{figure*}

\subsection{\textit{In and out-of-plane DFT-calculated band dispersions as a function of the CDW order parameter}}

At $T_{\text{CDW}}\approx 200$ K, 1$T$-TiSe$_2$ undergoes a phase transition towards a 2 $\times$ 2 $\times$ 2 commensurate inversion-symmetric CDW phase \cite{Hildebrand2018}. A weak PLD develops involving energetically favorable Ti-Se bond shortening of $\sfrac{3}{4}$ of the Ti atoms of the unit cell and a rotation of the Se atoms around the remaining $\sfrac{1}{4}$ \cite{whangbo1992, salvo1976} [Fig. S\ref{figS1} (a)]. It has been shown that the magnitude of the Ti-Se bond shortening can be associated with the CDW order parameter $\Delta$ as it follows a mean-field-like evolution as a function of temperature, characteristic of a second-order phase transition \cite{salvo1976}. In our DFT calculations we therefore vary the amplitude of the PLD in order to study the impact of the $\Delta$ reduction induced by K doping on the electronic band dispersions along the $ALA$ and $A \Gamma A$ directions of the 3D BZ [Fig. S\ref{figS1} (b)]. 

Fig. S\ref{figS1} (c) and S\ref{figS1} (d) respectively show the in-plane and out-of-plane DFT-calculated band dispersions of 1$T$-TiSe$_2$ of the non-distorted state (no PLD, red color scale) and fully distorted (100 $\%$ PLD, blue color scale) CDW phase. The color scale represents the distribution of spectral weight. The band structure of the fully hybridized CDW state have been shifted in energy in order to align the bottom of the Ti 3$d$ conduction bands with the one of the non-distorted state as its binding energy remains unaffected by the CDW phase transition (see main text). 

In very good agreement with experiment [Fig. 3(a) of the main text], we see on Fig. S\ref{figS1} (c) the appearance, upon the PLD, of the renormalization of the Ti 3$d$ electron band at the $k$-vectors of the $e$-$h$ band crossings \cite{Jaouen2018}, and of the backfolding of the two Se 4$p$ hole bands coming from the $\Gamma$ point [Se 4$p$ (1) and Se 4$p$ (2)]. From the calculated $\Gamma$-$A$ $k_z$ dispersions [Fig. S\ref{figS1} (d)], we see that the CDW phase transition is accompanied by an overall energy shift towards higher binding energy of the Se 4$p$ bands as well as a strong flattening of the topmost hole band [Se 4$p$ (1)], that becomes quasi two-dimensional (2D). Comparing the binding energies of the backfolded band extrema at $L$ Fig. S\ref{figS1} (c) with those at $\Gamma$ [Fig. S\ref{figS1} (d)], demonstrates that the heaviest Se 4$p$ band at $L$ [Se 4$p$ (2)] corresponds to the hole band at $\Gamma$ which does not play a role in the CDW ordering and has a Se 4$p_z$ character \cite{Watson2019}. The hole band of lower effective mass that appears at higher binding energy in Fig. S\ref{figS1} (c) [Se 4$p$ (1)] is associated with the 2D hole band of $p_{x,y}$ orbital character. Note that introducing the PLD in the DFT calculations that give a \textit{semimetallic normal state} with a significant $e$-$h$ band overlap ($\sim$170 meV) is sufficient to nicely match the experimental band dispersions of the CDW state.  

Fig. S\ref{figS1} (e) and S\ref{figS1} (f) focus on the in-plane and out-of-plane DFT band structures for slightly displaced atoms (25 $\%$ PLD, orange color scale), corresponding to the $\Delta$ reduction induced by the strongest K-doping of 0.19 $e^-$ per unit cell. Since the Ti 3$d$ band at $L$ rigidly shifts towards higher binding energy upon doping, the band structures have been shifted to take into account the chemical potential change of $\sim$180 meV after a first alignment of the bottom of the Ti 3$d$ conduction bands with the one of the non-distorted state. Overall, the in-plane DFT-calculated band dispersions [Fig. S\ref{figS1} (e)] very well reproduce the main spectroscopic features of the ARPES intensity map of the K-doped sample measured at $L$ using He-I radiation [Fig. 3(b) of the main text]; namely, the band binding energies, the effective masses, as well as the spectral weights. The backfolded hole bands [Se 4$p$ (1), (2)] are now nearly-degenerated at $L$, the spectral weight of the Ti 3$d$ band reduces approaching the $L$ point and the Mexican hat is almost not visible. On the contrary, the latter appears clearly in the DFT calculations at the $\Gamma$ point [green Se 4$p$ (1) in Fig. S\ref{figS1} (f)] as in the laser-based photoemission experiment [Fig. 3(g) and (h) of the main text]. Looking at the $k_z$ dispersion along $\Gamma$-$A$ Fig. S\ref{figS1} (f), finally allows to demonstrate that the two new hole bands with significant spectral weight appearing in the photoemission intensity map using 6.3 eV photon energy [Fig. 3(g) and (h) of the main text] are the Se 4$p_{x,y}$ valence bands derived from the $A$ point [black Se 4$p$ (1) and Se 4$p$ (2)]. Indeed, both the spectral weights of the hole bands at the $\Gamma$ and the $A$ points and the alternation of the band extrema [see the green, orange and grey circles on Fig. S\ref{figS1}(f) that respectively indicate the extrema of the $\Gamma$ and $A$-derived bands and Fig. 3(i) of the main text] well correspond to our laser-ARPES data. 

To summarize, our ARPES data, the simulated near-$E_F$ dispersions and the unfolded DFT band structure calculations are in excellent agreement and allow for demonstrating final-state effects of photoemission beyond the free-electron final state approximation \cite{Lindroos1996} in our laser-based photoemission data. Direct transitions from both the $\Gamma$ and the $A$-derived valence bands to Bloch-like final states allow for accessing the $\Gamma$-$A$ $k_z$ dispersion in the ARPES intensity map of the K-doped CDW state.

\end{document}